\documentclass[aps,prc,twocolumn,floatfix,superscriptaddress]{revtex4-1}  
\usepackage{amsmath}
\usepackage{amsfonts}
\usepackage{amssymb}
\usepackage{graphicx}
\usepackage{epstopdf}
\usepackage{xcolor}

\begin{document}
\title{Determination of asymptotic normalization coefficients for the channel $^{16}$O$\to \alpha+^{12}$C. Excited state 
	$^{16}$O($0^+; 6.05$ MeV)}

\author{L. D. Blokhintsev}
\affiliation{Skobeltsyn Institute of Nuclear Physics, Lomonosov Moscow State University,  Moscow 119991, Russia}

\author{A. S. Kadyrov}
\affiliation{Department of Physics and Astronomy and Curtin Institute for Computation, Curtin University, GPO Box U1987, Perth, WA 6845, Australia}

\author{A. M. Mukhamedzhanov}
\affiliation{Cyclotron Institute, Texas A\&M University, College Station, TX 77843, USA}

\author{D. A. Savin}
\affiliation{Skobeltsyn Institute of Nuclear Physics, Lomonosov Moscow State University,  Moscow 119991, Russia}

\begin{abstract}
Asymptotic normalization coefficients (ANC)  determine the overall normalization of cross sections of peripheral radiative  capture reactions.
In the present paper, we treat the ANC $C$ for the virtual decay $^{16}$O$(0^+; 6.05$ MeV)$\to \alpha+^{12}$C(g.s.), the known values of which are characterized by a large spread $(0.29-1.65)\times 10^3$ fm$^{-1/2}$.  The ANC 
$C$ is found by analytic continuation in energy of the $\alpha^{12}$C $s$-wave scattering amplitude, known from the phase-shift analysis of experimental data,  to the pole corresponding to the $^{16}$O bound state and lying in the unphysical region of negative energies. 
To determine $C$, two different methods of analytic continuation are used. In the first method,  the  scattering data are approximated by the sum of polynomials in energy in the physical region and then extrapolated to the pole. The best way of extrapolation is chosen on the basis of the exactly solvable model. Within the second approach, the ANC $C$ is found by solving the Schr\"odinger equation for the two-body $\alpha^{12}$C potential, the parameters of which are selected from the requirement of the best description of the phase-shift analysis data at a fixed experimental binding energy of $^{16}$O$(0^+; 6.05$ MeV) in the $\alpha+^{12}$C channel. The 
 values of the  ANC $C$ obtained within these two methods lie in the interval (886--1139) fm$^{-1/2}$. 
\end{abstract}

\maketitle

\section{Introduction}

Asymptotic normalization coefficients (ANC) determine the asymptotics of nuclear wave functions in binary channels at distances between fragments exceeding the radius of the nuclear interaction (see the recent review paper \cite{MBrev} and references therein). In terms of ANCs, the cross sections of peripheral nuclear processes are parameterized, such as reactions with charged particles at low energies, when, due to the Coulomb barrier, the reaction occurs at large distances between fragments. The most important class of such processes is astrophysical 
nuclear reactions occurring in the cores of stars, including the Sun. The important role of ANCs in nuclear astrophysics was first noted in Refs. \cite{Mukh1,Xu}, where it was shown that ANCs determine the overall normalization of cross sections of peripheral radiative  capture reactions (see also Refs. \cite{Mukh2,Mukh3}). 

We note that ANCs are important not only for astrophysics. ANCs turn out to be noticeably more sensitive to theoretical models than such quantities as binding energies or root-mean-square radii. This circumstance makes it possible to use a comparison of the
calculated and experimental ANC values to assess the quality of theoretical models. ANCs should be included in the
number of important nuclear characteristics along with such quantities as binding energies, probabilities of electromagnetic transitions, etc.

One of the most important astrophysical reactions is the radiative capture of $\alpha$ particles by $^{12}$C. The $^{12}$C$(\alpha,\gamma)^{16}$O reaction is activated during the helium burning stages of stellar evolution. It determines the relative abundance of $^{12}$C and $^{16}$O in the stellar core. 
Although the main contribution to the astrophysical factor of the $^{12}$C$(\alpha,\gamma)^{16}$O process at astrophysial energies comes from two subthreshold bound states $1^{-}$ and $2^{+}$, the radiative capture to 
the excited state $^{16}{\rm O}(0^+; 6.05$ MeV) also contributes. Owing to the small binding energy of the bound state  $(0^+; 6.05 {\rm MeV})$, the $E1$ 
transition $^{12}{\rm C}(\alpha,\gamma)^{16}{\rm O}(0^+; 6.05 {\rm MeV})$ to this state at lower energies relevant the radiative capture  is peripheral. The normalization of the astrophysical $S$-factor for this transition is determined by the ANC for the virtual decay $^{16}$O$^*\to \alpha+^{12}$C(g.s.), where g.s. stands for the ground state.
 Hence the  knowledge of this ANC is important. 
 
	However, the available in literature ANC values  for the channel $^{16}{\rm O}(0^+; 6.05 {\rm MeV}) \to \alpha+^{12}$C(g.s.)  obtained by various methods are characterized by a noticeable  spread (see Table \ref{table1}).  In this paper, we determine the ANC for this channel using analytic continuation in the energy plane
 of the $\alpha^{12}$C $s$-wave scattering amplitude, known from the phase-shift analysis of experimental data. 
 Since we use the analytic continuation, one may consider the obtained value as an experimental one. 
 
In what follows, the ANC for this channel will be referred to as $C$. The binding energy corresponding to the virtual decay $^{16}$O$(0^+; 6.05$ MeV)$\to \alpha+^{12}$C(g.s.) is $\varepsilon=1.113$ MeV.
 
 The value of ANC $C$ is determined by analytical continuation in center of mass (c.m.) energy $E$ of the  partial $S$-wave amplitude $f_0(E)$ of elastic scattering of alpha particles on $^{12}$C to a point corresponding to the excited  $^{16}$O$(0^+)$ bound state  and lying in the unphysical region of negative  values of $E$. Information on $f_0(E)$ at $E>0$ is taken from the phase-shift analysis. Various methods of analytic continuation are used. The obtained ANC values are compared with the results of other authors.

The paper is organized as follows. Section II presents the general formalism of the method used. Section III is devoted to the choice of the best method to continue the experimental data within the exactly solvable model. Determining ANC $C$ from the analytic continuation of the phase-shift analysis data is outlined in Section IV.  The results are discussed in Section V. 

We use the system of units in which $\hbar=c=$1 throughout the paper.

\begin{table}[htb]
	\caption{ANC $C$ values for $^{16}$O$(0^+; 6.05$ MeV)$\to \alpha+^{12}$C(g.s.).}
	\begin{center}
		\begin{tabular}{ccc}
			\hline \hline
			$C$, fm$^{-1/2}$ &&  Reference \\
			\hline
			$(1.56\pm0.09)\times10^3$ &&  [6] \\
			$0.406\times10^3$  &&  [7] \\
			$(0.64-0.74)\times10^3$  &&  [8] \\
			$0.293\times10^3$ &&  [9] \\
			\hline \hline
		\end{tabular}
	\end{center}
	\label{table1}
\end{table}

\section{Basic formalism}

In this section we recapitulate basic formulas which are necessary for the subsequent discussion. 

The Coulomb-nuclear  amplitude of elastic scattering of particles 1 and 2 is of the form
\begin{equation}\label{fNC}
f_{NC}({\rm {\bf  k}})=\sum_{l=0}^\infty(2l+1)\exp(2i\sigma_l)\frac{\exp(2i\delta_l)-1}{2ik}P_l(\cos\theta).
\end{equation}
Here ${\rm {\bf k}}$ is the relative momentum of particles 1 and 2, $\theta$ is the c.m. scattering angle,   
$\sigma_l=\arg\,\Gamma(l+1+i\eta)$  
 and $\delta_l$ are the pure Coulomb and Coulomb-nuclear phase shifts, respectively, $\Gamma(z)$ is the Gamma function, 
\begin{equation}\label{eta}
\eta =Z_1Z_2e^2\mu/k
\end{equation}
is the Coulomb  parameter for the 1+2 scattering state with the relative momentum $k$ related to the energy by  $k=\sqrt{2\mu E}$,
$\mu=m_1m_2/(m_1+m_2)$, $m_i$ and $Z_ie$  are the mass and the electric charge of particle $i$.

The behavior of the Coulomb-nuclear partial-wave amplitude $f_l=(\exp(2i\delta_l)-1)/2ik$ is irregular near 
$E=0$. Therefore, one has to introduce the renormalized Coulomb-nuclear partial-wave amplitude $\tilde f_l$ \cite{Hamilton,BMS,Konig}
\begin{equation}\label{renorm}
\tilde f_l=\exp(2i\sigma_l)\,\frac{\exp(2i\delta_l)-1}{2ik}\,\left[\frac{l!}{\Gamma(l+1+i\eta)}\right]^2e^{\pi\eta}.
\end{equation}
Eq.~(\ref{renorm}) can be rewritten as 
\begin{equation}\label{renorm1}
\tilde f_l=\frac{\exp(2i\delta_l)-1}{2ik}C_l^{-2}(\eta),
\end{equation}
where $C_l(\eta)$ is the Coulomb penetration factor (or Gamow factor) determined by
\begin{align}\label{C}
C_l(\eta)&=\left[\frac{2\pi\eta}{\exp(2\pi\eta)-1}v_l(\eta)\right]^{1/2}, \\ 
v_l(\eta)&=\prod_{n=1}^{l}(1+\eta^2/n^2)\;(l>0),\quad v_0(\eta)=1.
\end{align}
It was shown in Ref. \cite{Hamilton} that  the
analytic properties of ${\tilde f}_{l}$ on the physical sheet of $E$  are analogous to the ones of the partial-wave scattering amplitude for the short-range potential and ${\tilde f}_{l}$  can be analytically continued into the negative-energy region.

The amplitude $\tilde f_l$ can be expressed in terms of the Coulomb-modified effective-range function (ERF) $K_l(E)$ \cite{Hamilton, Konig}  as
\begin{align} 
\label{fK}
\tilde f_l&=\frac{k^{2l}}{K_l(E)-2\eta k^{2l+1}h(\eta)v_l(\eta)}\\ 
&=\frac{k^{2l}}{k^{2l+1}C_l^2(\eta)(\cot\delta_l-i)} \\ 
&=\frac{k^{2l}}{v_l^2 k^{2l}\Delta_l(E)-ik^{2l+1}C_l^2(\eta)},
\label{fK3}
\end{align}   
where
\begin{align}\label{scatfun}
K_l(E)&= k^{2l+1} \left[ C_l^2(\eta)(\cot\delta_l-i) + 2 \eta h(k)v_l(\eta) \right],\\ 
h(\eta) &= \psi(i\eta) + \frac{1}{2i\eta}-\ln(i\eta), \\  
\Delta_l(E)&=kC_0^2(\eta)\cot\delta_l, 
\label{Deltal}
\end{align}
$\psi(x)$ is the digamma function and $\Delta_l(E)$ is the $\Delta$ function introduced in Ref. \cite{Sparen}. 

If the $1+2$ system has in the partial wave $l$ the bound state 3 with the binding energy $\varepsilon=\varkappa^2/2\mu>0$, then the amplitude $\tilde f_l$ has a pole at $E=-\varepsilon$. The residue of $\tilde f_l$ at this point is expressed in terms of the ANC
$C^{(l)}_{3\to 1+2}$ \cite{BMS} as
\begin{align}\label{res2}
{\rm res}\tilde f_l(E)|_{E=-\varepsilon}&=\lim_{\substack{E\to -\varepsilon}}[(E+\varepsilon)\tilde f_l(E)] \\
&=
-\frac{1}{2\mu}\left[\frac{l!}{\Gamma(l+1+\eta_b)}\right]^2 \left[C^{(l)}_{3\to 1+2}\right]^2,
\label{res22}
\end{align}
where $\eta_b=Z_1Z_2e^2\mu/\varkappa$ is the Coulomb  parameter for the bound state 3.

Formally, the most natural quantity for continuing the scattering data to the region of negative energies is the ERF $K_l(E)$ which is
expressed in terms of scattering phase shifts.  It was shown in Ref. \cite{Hamilton} that function $K_l(E)$ defined by (\ref{scatfun}) is analytic near $E=0$ and can be expanded into a Taylor series in $E$. In the absence of the Coulomb interaction ($\eta=0$), $K_l(E)=k^{2l+1}\cot\delta_l(k)$. However, in case of charged particles, 
the ERF for the short-range interaction should be modified. Such modification 
generates additional terms in the ERF (see Eq. (\ref{scatfun})). These terms depend only on the Coulomb interaction and may far exceed, in the absolute value, the informative part of the ERF containing the phase shifts. This fact may hamper the practical procedure of the analytic continuation and affect its accuracy. In particular, for the $\alpha+^{12}$C system  considered in this paper, any reliable continuation of  $K_0(E)$ to the region $E<0$, taking into account experimental errors, turned out to be impossible.
It was suggested in Ref.~\cite{Sparen} to use for the analytic continuation the quantity $\Delta_l(E)$ rather than the ERF $K_l(E)$. The
$\Delta_l(E)$ function does not contain the pure Coulomb terms. 

In what follows, for the analytical continuation of the experimental data, we will use the function $\Delta_l(E)$ at $l=0$ and various analytic expressions composed of it ($\Delta$-method). Within this method, the real part of the denominator of the amplitude 
$\tilde f_0(E)$, which for $E > 0$ coincides with $\Delta_0(E)$ (see (9)), is approximated by polynomials in $E$ and continued analytically to the region $E < 0$. The amplitude pole condition is formulated as $\Delta_0^{appr}(-\varepsilon)=0$, where $\Delta_0^{appr}(E)$ is a function approximating $\Delta_0(E)$ at $E>0$. From the results of Refs. \cite{BKMS2,Gaspard}
it follows that the $\Delta$-method, although non-strict and approximate, is sufficiently accurate for the system under consideration and the energy range of interest. Note that for lighter systems, in particular for the channels $^6$Li$\to \alpha+d$ and  
$^7$Be$\to \alpha+^3$He,  the $\Delta$-method is not suitable.

The functions we are considering, determined by the experimental data, are approximated in the physical region $E>0$ by the expression 
\begin{equation}\label{polin} 
\sum_{i=0}^Nc_i P_i(E), 
\end{equation}
where $P_i$ are the Chebyshev polynomials of degree $i$. The maximum degree of the polynomial $N$ and the coefficients $c_i$ are determined from the best description of the approximated functions using the $\chi^2$ criterion and also the $F$-criterion (see the monograph 
\cite{Wolberg}). Note that these criteria give similar results.

\section{Model analysis to choose the best option to continue experimental data}

In this section, within the framework of an exactly solvable model, a comparative analysis of various methods of continuing the scattering data to the pole point of the partial-wave scattering amplitude is carried out to choose the best way of determining the  ANC. The experimental values of phase shifts are simulated by the results of calculations in a two-particle model with a potential taken in the form of a square well plus the Coulomb interaction. To the authors' knowledge, the square-well potential is the only local potential which, with the added Coulomb interaction, permits the analytic solution of the Schr\"odinger equation at any value of the orbital angular momentum $l$. The two parameters of the square-well potential, the radius $R$  and the depth $V_0$  were adjusted to reproduce, in the presence of two bound $0^+$  states, the  experimental binding energy of the upper state $\varepsilon=1.113$ MeV and the ANC value $C=690.0$ fm$^{-1/2}$, which is the average value obtained in Ref. \cite{Ando}. The calculations in this section are methodological, and the qualitative conclusions obtained should not depend on the choice of a specific ANC value within the values presented in Table~\ref{table1}.

Solving the Schr\"odinger equation within the aforementioned model results in the following expression for the phase shift $\delta_l$ 
\cite{BKMS1}
\begin{align}
\label{cotdelta}
\cot\delta_l  & \nonumber \\
=&\dfrac{\dfrac{d\hat G_{l,\eta}(k,R)}{dR} \hat F_{l,\eta_1}(K,R)
- \dfrac{d\hat F_{l,\eta_1}(K,R)}{dR} \hat G_{l,\eta}(k,R)} 
{\dfrac{d\hat F_{l,\eta}(k,R)}{dR} \hat F_{l,\eta_1}(K,R)
- \dfrac{d\hat F_{l,\eta_1}(K,R)}{dR} \hat F_{l,\eta}(k,R)} .
\end{align}
Here $K=\sqrt{2\mu(E+V_0)}$, $\hat F_{l,\eta}(q,r)= F_l(\eta, qr)/qr$, $\hat G_{l,\eta}(q,r)= -G_l(\eta, qr)/qr$, $F_l(\eta, \rho)$ and 
$G_l(\eta, \rho)$ are the regular and irregular Coulomb functions, respectively~\cite{NIST}.
Eq.(\ref{cotdelta}) allows one to calculate the function $\Delta_l(E)$ using Eqs. (5) and (12).

For the model phase-shift analysis, 39 points in the c.m. energy $E$ were taken in the range 1.47--6.56 MeV, which is close to the range 1.96--4.97 MeV, for which phase shifts were obtained in Ref. \cite{Tischhauser} from the analysis of experimental data. The theoretical phase shifts calculated at these points, as in \cite{Tischhauser}, were superimposed with a random error of 5\%.
 
To approximate the function $\Delta_0(E)$ for $E>0$ and extend it to the point $E=-\varepsilon$, four different ways (versions) were chosen:
\begin{itemize}
	\item[]  Version 1 -- continuation of the function $\Delta_0(E)$ directly, 
	\item[] Version 2 -- continuation of the function $\dfrac{\Delta_0(E)}{E+\varepsilon}$, 
	\item[] Version 3 -- continuation of the function $\ln(A- \Delta_0(E))$, 
	\item[] Version 4 -- continuation of the function
	$\ln \left(\dfrac{-\Delta_0(E)}{E+\varepsilon}\right)$. 
\end{itemize}
The appearance of the ln sign in Versions 3 and 4 is due to the fact that near $E=0$,  $\Delta_0(E)$ changes exponentially; using the logarithmic function makes it possible to soften this dependence and improve the quality of approximation of the considered functions by polynomials. The constant $A>0$ is added to make $A-\Delta_0(-\varepsilon)$ positive. Note that in the energy range under consideration, $\Delta_0(E)<0$ and decreases monotonically as $E$ increases; for 
$E\to -\varepsilon$, $\Delta_0(E)\to 0$. The value of $A$ is chosen so that the condition
$A\ll|\Delta_0(E)|$ holds, and the approximated function is as close to a straight line as possible so that it could be approximated by a polynomial of a low degree. Under these conditions, the calculation results are little sensitive to changes in $A$.

Note that within Versions 2 and 4 the condition $\Delta_0^{appr}(-\varepsilon)=0$ is met automatically. In Versions 1 and 3, the fulfillment of this condition with high accuracy is achieved by the fact that the point $E=-\varepsilon$ is included in the set of points used in the approximation of the corresponding functions, and the error at this point is taken to be many orders of magnitude smaller than 5\% corresponding to the points at $E>0$.

The $C$ values obtained in Versions 1--4 are compared with the exact value $C=690.0$ fm$^{-1/2}$ for the chosen potential. 
It follows from the calculation results 
that the closest to the exact value of $C$, as well as the best convergence of the results with an increase in the maximum degree of approximating polynomials $N$, correspond to Version 3.

\section{Finding ANC $C$ from phase-shift analysis data}

First, ANC $C$ is found directly by continuing to the pole $E=-\varepsilon$ phase shifts obtained from the phase-shift analysis of the elastic $\alpha-^{12}$C scattering data of Ref. \cite{Tischhauser}.
   For fitting, 20 points are used for the laboratory energy $E_\alpha$ in the range 2.607 - 6.620 MeV (a narrow resonance is higher in energy).
Based on the results of the previous section, 
we use Version 3 -- the continuation of the function $\ln(A-\Delta_0(E))$ as the most stable one. 
Within this version, to determine the sensitivity of the results to parameter $A$, calculations have been performed for two different 
$A$ values: $A_1=0.506\times 10^{-5}$ fm$^{-1}$ and $A_2=0.805\times 10^{-5}$ fm$^{-1}$.
 Using the $\chi^2$ and $F$ criteria,  we obtain $C=1175$ fm$^{-1/2}$ and $C=1097$ fm$^{-1/2}$ for $A_1$ and $A_2$, respectively. It can be seen that these two values are close to each other.
Calculations of $C$ were also carried out using 10 experimental points lying in a narrower energy interval
(up to $E_\alpha=4.31$ MeV). In this case, $C=1139$ fm$^{-1/2}$ is obtained. This ANC value lies between two values obtained over a wider energy range.

Next, we use a different approach to determine the ANC $C$ based on the phase-shift analysis from Ref.  \cite{Tischhauser}. The approach is based on fitting parameters of a potential. The square-well potential parameters are selected by the $\chi^2$ method from the requirement of the best description of the phase-shift analysis data at a fixed experimental binding energy of $\varepsilon=1.113$ MeV. After that, ANC is found from the solution of the Schr\"odinger equation for the square well with the established parameters plus the Coulomb interaction. Such an approach can be formally considered as an alternative way of analytical continuation of the scattering data. The square well  with both two and three bound states was considered. Wide and narrow energy ranges were used for fitting. At the same time, it was also checked how accurately the square-well potential describes the data of the phase-shift analysis with parameters adjusted by the value $\varepsilon=1.113$ MeV and the ANC values previously obtained by the other authors and presented in Table \ref{table1}.

\begin{figure}[htb]
\includegraphics[scale=1.0]{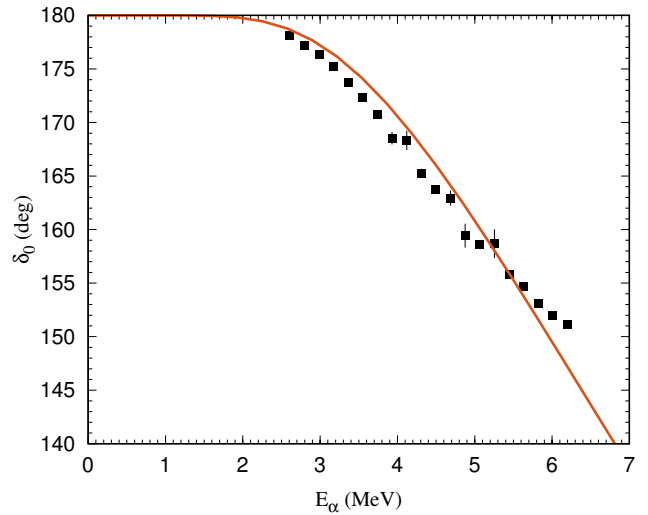} 
\caption {Phase shift $\delta_0$ for $\alpha^{12}$C scattering (wide energy range). Solid line corresponds to the square-well potential and ANC $C=0.734\times 10^{3}$ fm$^{-1/2}$. Experimental points are taken from Ref.  \cite{Tischhauser}.}
\label{figx1}
\end{figure}

The results for $C$ and $\chi^2$, obtained using a wider energy interval and a two bound-state potential, are shown in Table \ref{table2}. The best result for $\chi^2$ corresponds to $C=734$ fm$^{-1/2}$.  Parameters of the potential are $V_0$=25.7656 MeV and $R$=3.81962 fm. Figure \ref{figx1} shows 
phase shift $\delta_0$ for $\alpha^{12}$C scattering obtained using the wide energy range. One can see
that near the upper boundary of the considered energy range, the calculated phase shift begins to deviate from the results of the phase-shift analysis. This suggests that the square-well potential cannot accurately describe such a wide energy range. Therefore, a similar fitting was carried out for a narrower interval, which was already used in Section III. 

\begin{table}[htb]
\caption{ANC $C$  for $\alpha+^{12}$C (wide energy range).}
\begin{center}
\begin{tabular}{ccc}
\hline \hline 
$C$, fm$^{-1/2}$ && $\chi^2$ \\
\hline
0.780$\times 10^{3}$ && 347.4 \\
0.734$\times 10^{3}$ && 175.6 \\
0.732$\times 10^{3}$ && 175.9 \\
0.730$\times 10^{3}$ && 176.5 \\
\hline \hline
\end{tabular}
\end{center}
\label{table2}
\end{table}

\begin{figure}[htb]
\includegraphics[scale=1.0]{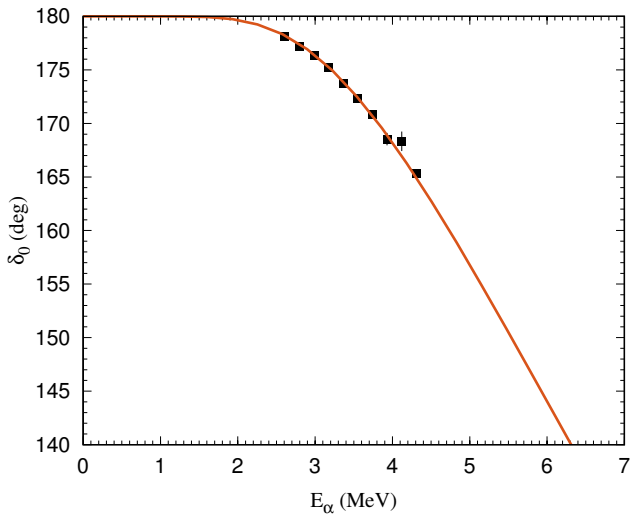} 
\caption {The same as in Fig. \ref{figx1} but for the narrow energy range.  ANC $C= 0.938\times 10^{3}$  fm$^{-1/2}$.}
\label{figx2}
\end{figure}

 The results are presented in Table \ref{table3} and in Fig. \ref{figx2}. The best result for $\chi^2$ corresponds to $C=938$ fm$^{ -1/2}$. Parameters of the corresponding potential are $V_0$=22.7495 MeV and $R$= 4.16411 fm. Note that $\chi^2$ for a narrow energy range is more than two orders of magnitude less than for the wide interval and is close to unity. The best agreement is also seen in the figure. Therefore, the narrow interval should be assessed as more adequate, and the results obtained for it are closer to the physical ones.
 
 For comparison, the analogous calculations were performed for the narrow range for the square-well potential with three bound states as well.  In this case, the best result is $C=886$ fm$^{-1/2}$, which is close to the value 938 fm$^{-1/2}$ obtained for the two bound state case.

\begin{figure}[htb]
\includegraphics[scale=1.0]{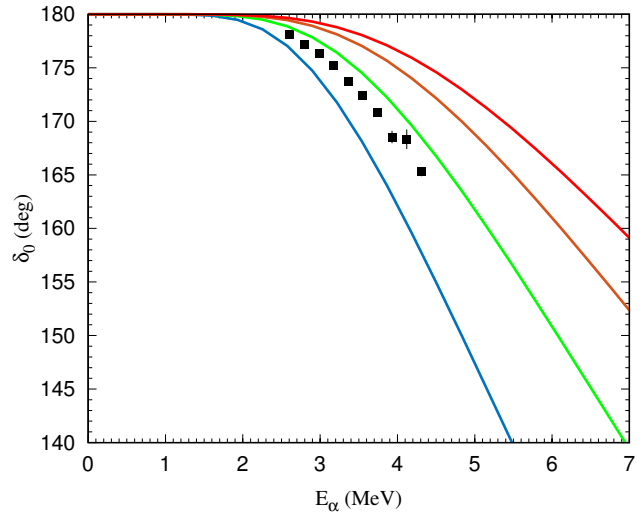} 
\caption {Phase shift $\delta_0$ for $\alpha^{12}$C scattering (narrow energy range). Solid lines correspond to the square-well potential with ANC $C$ taken from Table \ref{table1}. Blue line: $C=1.56\times 10^{3}$ fm$^{-1/2}$ \cite{Avila}; green line: $C=0.690\times 10^{3}$ fm$^{-1/2}$  \cite{Ando}; dark red line: $C=0.406\times 10^{3}$ fm$^{-1/2}$ \cite{Orlov2}; red line: $C=0.293\times 10^{3}$ fm$^{-1/2}$ \cite{Orlov3}. Experimental points are taken from Ref. \cite{Tischhauser}.}
\label{figx3}
\end{figure}

Phase shift calculations were also carried out for the square-well potential with two bound states and parameters adjusted to $\varepsilon=
1.113$ MeV and the ANC values obtained by the other authors and listed in Table \ref{table1}. The corresponding results are presented in Table \ref{table4} and in Fig. 
\ref{figx3}.
 
\begin{table}[htb]
\caption{ANC $C$  for $\alpha+^{12}$C (narrow energy range).}
\begin{center}
\begin{tabular}{ccc}
\hline \hline
$C$, fm$^{-1/2}$ && $\chi^2$ \\
\hline
0.899$\times 10^{3}$ && 6.2831 \\
0.938$\times 10^{3}$ && 0.7756 \\
0.939$\times 10^{3}$ && 0.7764  \\
0.972$\times 10^{3}$ && 4.2636 \\
\hline \hline
\end{tabular}
\end{center}
\label{table3}
\end{table}

\begin{table}[htb]
\caption{ANC $C$  for $\alpha+^{12}$C (narrow energy range).}
\begin{center}
\begin{tabular}{ccccc}
\hline \hline
$C$, fm$^{-1/2}$ && $\chi^2$ &&  Reference \\
\hline
1.560$\times 10^{3}$ &&   410 && \cite{Avila} \\
0.690$\times 10^{3}$ &&   409 &&  \cite{Ando} \\
0.406$\times 10^{3}$ &&  6883 &&  \cite{Orlov2} \\
0.293$\times 10^{3}$ && 23840 &&  \cite{Orlov3} \\
\hline \hline
\end{tabular}
\end{center}
\label{table4}
\end{table}

\section{Conclusions}

In the present paper, we treated the ANC $C$ corresponding to the virtual decay $^{16}$O$(0^+; 6.05$ MeV)$\to \alpha+^{12}$C, the  values of which obtained by various methods are characterized by a large spread.  
To determine $C$, we use two different methods of analytic contiunuation in energy of experimental $\alpha-^{12}$C scattering data to the pole corresponding to the bound state $^{16}$O$(0^+; 6.05$ MeV). In the first method,
the function $\Delta_0(E)$ introduced in Ref. \cite{Sparen} and defined above in Eq. \eqref{Deltal} is approximated by the sum of the Chebyshev polynomials in the physical region $E>0$ and then extrapolated to the pole. The best way of extrapolation is chosen on the basis of the exactly solvable model.
Within the second approach,
the ANC $C$ is found by solving the Schr\"odinger equation for the square-well nuclear potential, the parameters of which are selected by the $\chi^2$ method from the requirement of the best description of the phase-shift analysis data at a fixed experimental binding energy of 
$^{16}$O$(0^+; 6.05$ MeV) in the $\alpha+^{12}$C channel. In both methods, wider and narrower energy ranges were used to adjust the parameters that determine the analytic continuation.
	If, in accordance with the results of Section IV, we assume that for the second method it is better to restrict ourselves to the data within the narrower energy range, then we can conclude that all the results obtained by us for ANC $C$ lie in the interval 
	(886--1139) fm$^{-1/2}$. If we take into account the data
	within the wider energy range, then the lower limit for $C$ is 
	734 fm$^{-1/2}$.
	
	 In connection with the use of the $\Delta$-method in this work, it should be emphasized that, within the framework of this method, it is not the function $\Delta_l(E)$ that actually is continued into the region of negative energies, but the real part of the denominator of the Coulomb-modified amplitude $\tilde f_l(E)$ defined in Eq. \eqref{fK3}. As we mentioned earlier, $\Delta_l(E)$ cannot be directly continued to the region $E < 0$ by means of polynomial approximation, since it has an essential singularity at $E=0$. For the sake of brevity, let us prove this assertion for $l=0$, although the following arguments are valid for arbitrary values of $l$. In accordance with Eq. \eqref{fK3}, $\tilde f_0(E)$ can be written as $\tilde f_0(E)=D_0^{-1}(E)$, 
where $D_0(E)=\Delta_0(E)-ikC_0^2(\eta)$. The function $C_0^2(\eta)$ defined in Eq. \eqref{C} possesses an essential singularity at $E=0$	due to the presence of $\exp(2\pi\eta)$ with $\eta=Z_1Z_2e^2\sqrt{\mu/2E}$ (see Eq. \eqref{eta}). On the other hand, $D_0(E)$ has no essential singularity at $E=0$ since the analytic properties of ${\tilde f}_{l}(E)$ on the physical sheet of $E$  are analogous to the ones of the partial-wave scattering amplitude for the short-range potential \cite{Hamilton}. Therefore, in the expression for $D_0(E)$, the essential singularity of the term $ikC_0^2(\eta)$ must be compensated by the essential singularity of $\Delta_0(E)$. In Ref.  \cite{BKMS1}, within the framework of an exactly solvable model, it is shown explicitly that  functions $ikC_0^2(\eta)$ and $\Delta_0(E)$ have essential singularities at $E=0$ and behave irregularly at 
$E\to-0$, but these irregularities are compensated in the expression for $D_0(E)$. From the above-stated it clearly follows that the statement about the absence of an essential singularity of $\Delta_l(E)$ at $E=0$, made in Refs.  \cite{Orlov3,Orlov4} is erroneous.
	
	In this work, we dealt with ANC for the channel $^{16}$O$(0^+; 6.05$ MeV)$\to \alpha+^{12}$C. Work to determine similar ANCs for excited states of $^{16}$O with $l>0$ is in progress. As for the ground state of $^{16}$O, it is hardly possible to determine the corresponding ANC  by analytic continuation of the data on partial-wave scattering amplitudes. As follows from the results of Refs. \cite{BKMS2,BlSav2016}, in the case when there is more than one bound state with the same quantum numbers in the system, the method of analytic extrapolation makes it possible to obtain reliable information only about the upper (weakest bound) state. 

\section*{Acknowledgements}

This work was supported by the Russian Foundation
for Basic Research Grant No. 19-02-00014 (L.D.B. and D.A.S.). A.S.K. acknowledges the support from the Australian Research Council. A.M.M. acknowledges the support from the US DOE National Nuclear Security Administration under Award Number DENA0003841 and DOE Grant No. DE-FG02-93ER40773.

\end{document}